# Intercalation of graphene on SiC(0001) via ion-implantation

A. Stöhr[1], S. Forti[1], S. Link[1], A.A. Zakharov[2], K. Kern[1,3], U. Starke[1], H.M. Benia[1,*]

[1]*Max-Planck-Institut für Festkörperforschung, 70569 Stuttgart, Germany*
[2]*MAXLab, Lund University, P.O. Box 118, Lund, S-22100, Sweden*
[3]*Institut de Physique de la Matière Condensée, Ecole Polytechnique Fédérale de Lausanne, 1015 Lausanne, Switzerland*

**Abstract:** Electronic devices based on graphene technology are catching on rapidly and the ability to engineer graphene properties at the nanoscale is becoming, more than ever, indispensable. Here, we present a new procedure of graphene functionalization on SiC(0001) that paves the way towards the fabrication of complex graphene electronic chips. The procedure resides on the well-known ion-implantation technique. The efficiency of the working principle is demonstrated by the intercalation of the epitaxial graphene layer on SiC(0001) with Bi atoms, which was not possible following standard procedures. Our results put forward the ion-beam lithography to nanostructure and functionalize desired graphene chips.

**Keywords:** Graphene, Intercalation, SiC, Ion-implantation, Bismuth

**Introduction:**

During the last decade graphene has become one of the most widespread and extensively investigated materials. Owing to its robust atomically thin carbon structure which hosts highly mobile electrons, graphene promises the upcoming of many novel and modern technological applications[1–4]. The key to graphene's various effects and uses resides in the ability to fine-tune its electronic and chemical properties, a goal that numerous previous and ongoing studies are focused on. In this regard, the functionalization of graphene has shown incontestable efficiency in inducing new and stable properties[5–9]. Besides, the correct selection of a support for the graphene that facilitates exploiting its properties is of great importance. The wide bandgap semiconductor SiC constitutes an excellent substrate. On its (0001) surface, high-quality epitaxial graphene (EG) can be grown in a large-scale, functionalized, and introduced at an advanced industrial production level[10–14]. The functional entity, the size of which may vary from a single atom to a macromolecule, can be introduced into the EG/SiC system either by adsorption, carbon substitution, or intercalation[11,15–17]. The latter seems in particular advantageous as an on-top arrangement of the graphene layer protects the intercalant. Recently, evidence for superconductive EG has been observed after Li adsorption[18], while controlled intercalation of Ge enables the fabrication of ballistic bipolar-junctions[19]. Nevertheless, one of the significant advances would be the ability to control spin-orbit coupling (SOC) effects in EG. It has been predicted that introduction of specific atomic impurities could enhance the SOC[20–24]. While it was argued that atoms with *d* valence electrons induce a substantial energy gap at the Dirac point and may turn the graphene into a 2D topological insulator, only for very few heavy metal adatoms the induction of a topological phase was anticipated[21–23,25,26]. The situation where Bi is introduced in an intercalation configuration is expected to induce a Rashba-like spin-splitting of the Dirac cone[24]. Yet so far, the intercalation of Bi in the EG/SiC system seemed not possible and the question whether it experimentally causes SOC enhancement remained unresolved. Only functionalization by transfer doping from a Bi adlayer was demonstrated[17]. Here, using a new experimental procedure, we could intercalate Bi under the EG and show that the resulting band structure does not present a band spin-splitting. Bi intercalation has been made possible by means of ion-implantation. Using a commercial e-beam evaporator we implanted energetic Bi ions through the so-called zero-layer graphene (ZLG)[10,27] and decoupled it from the SiC(0001) to form a quasi-free standing EG. We foresee the use of ion-implantation not only for the intercalation of materials that are still difficult to intercalate on SiC(0001), such as Ni or Sb, but also, for fabrication of advanced graphene nanostructures using lithography based on the focused ion-beam technique[27].

---

[*] h.benia@fkf.mpg.de

**Experimental details:**

On-axis oriented single crystalline 6H-SiC(0001) n-type samples purchased from SiCrystal GmbH were used in this study. The ZLG develops on the SiC(0001) surface after annealing in argon atmosphere at 1450 °C for 10 minutes. Details on the ZLG formation can be found elswhere[10,11]. The X-ray photoemission spectroscopy (XPS) and the angle-resolved photoemission spectroscopy (ARPES) experiments were performed at the MPI-Stuttgart using a hemispherical SPECS Phoibos 150 analyzer. The used excitation sources are Mg Kα (1253.6 eV) for the XPS and Helium II (40.8 eV) for the ARPES. LEEM, μ-LEED, and μ-ARPES were performed at the MAXlab synchrotron radiation facility in Lund (Sweden). The spin-resolved ARPES measurements (SARPES) were carried out with a Phoibos hemispherical energy analyzer equipped with a Mott spin polarimeter at BESSY-II (Helmholtz-Zentrum Berlin).

To intercalate Bi under graphene on SiC(0001) we have first tried to follow a procedure which leads to successful intercalation of other materials like Au and Ge[11,28,29]. This procedure consists of material evaporation on the ZLG at room temperature then annealing at higher temperatures. In the case of Bi, such a method does not lead to the decoupling of the ZLG and the deposited Bi evaporates at a temperature around 500 °C from the surface and vanishes completely at 600 °C. We alternatively use low-energy ion-implantation in the SiC surface using an electron beam evaporator. In standard e-beam evaporators the positive ions created by the collision of the electron beam with evaporated atoms can be accelerated to high energies. Most of the created ions are collected by the perforated Faraday-cup used to monitor the evaporation flux and only approximately 3 % pass through. On our 5×7 mm² SiC samples we get Bi ion-currents ranging from 30 nA to 50 nA for a flux of 2 to 3 µA at the Faraday-cup. Bi ions are accelerated onto the sample by the voltage applied to the crucible. Here, we used 550 V. To avoid having Bi material on top the ZLG, the samples were heated during the implantation by direct-current-heating to 700 °C, which is higher than the Bi sublimation point (~520 °C). Using the SRIM program[30] we calculated the Bi implantation depth as well as the concentration of the vacancies created by the energetic impinging Bi ions on SiC (without the ZLG). From the calculation results shown in figure 1.a, Bi ions lose their energy at a depth of about 35 Å from the surface. Along their way within the SiC sample they damage the surface by sputtering and vacancy creation. The calculated average sputtering yield for C and Si atoms is 0.03 and 0.08, respectively. The created vacancies are concentrated at a depth around 6 Å (Figure 1.a).

**Results and discussion:**

Experimentally, the damage caused by the impinging ions can be observed in figure 1.b to 1.e. Prior to implantation, the measured band structure in the vicinity of the $\bar{K}$-point and the C 1s core level spectrum show clear signatures of a clean ZLG[10]. The two nondispersive states at $E_B = 0.45$ eV and $E_B = 1.6$ eV as well as the core levels S1 ($E_B = 284.75$ eV) and S2 ($E_B = 285.5$ eV) are consequences of ZLG formation on the SiC(0001) surface[10]. After exposure to the Bi beam, the signal of the nondispersive states is clearly reduced in the ARPES data (figure 1.c) and the intensity of the S1 and S2 peaks weakens dramatically in the XPS spectrum (figure 1.e). Furthermore, two new signals appear at $E_B = 284.6$ eV and $E_B = 282.8$ eV which we attribute to patches of decoupled graphene[10] and defects induced in the SiC substrate[31], respectively (figure 1.e). The signal at $E_B = 285.63$ eV, which is shifted by more than +0.1 eV from S2, is also ascribed to the presence of defects. Hence, the energetic Bi ions cause damage to the structure of the ZLG, which results in patches with differently bonded carbon[31]. At the same time, the XPS spectra in figure 1.f show that Bi is effectively present within the surface of the sample despite the high sample temperature during the implantation. The Bi 4f spectrum has two peaks at different binding energies and with different intensities. The peak at higher binding energy with relatively low intensity is attributed to implanted Bi atoms in the bulk, since its intensity diminishes when the detection is more restricted to photoelectrons emitted from the surface. The presence of Bi at the top layers is attributed to diffusion towards the surface enabled by the high sample temperature. The Bi 4f photoemission spectrum remains stable even after exposure to air and after the various treatments discussed below, where the sample is heated at high temperatures in different atmospheres. This suggests that Bi is strongly bound to the surface and in addition protected against interaction with oxygen and moisture present in air by the overlaying carbon layer.

To reduce the defects and restructure the surface, we chose first to anneal the samples in vacuum. Heating to 900 °C for 1 hour results in the appearance of π-bands of *n*-doped graphene without the nondispersive bands that characterize the ZLG (see figure 2.a). This indicates a successful decoupling of graphene using Bi implantation. First-principles calculations predict that graphene intercalation with Bi on SiC(0001) induces a similar *n*-doping[24]. However the annealing procedure described above does not produce high quality graphene as indicated by broad and weak π-bands (figure 2.c). A band width of $\Delta k = 0.14$ Å$^{-1}$ ± 0.03 Å$^{-1}$, which is more than twice as large as for a monolayer-graphene (MLG)[32], was extracted after the fit of the momentum distribution curve (MDC) at $E_B = 1$ eV with Lorentzian curves. Prolonged annealing at 900 °C does not lead to any noticeable changes in the band structure. This means the energy barrier for removing the defects is not reached with 900 °C and one has to increase the temperature. Yet, increasing the annealing temperature in UHV causes surface graphitization[33]. Nevertheless, since the onset-temperature for Si depletion increases with vapor-pressure, we tested heating the sample in Argon and in methane atmospheres. In the case of Argon gas, after Bi-implantation, the sample is transferred through ambient conditions to a quartz glass reactor. After sample outgassing in UHV conditions, the reactor is filled with 1 bar Ar. The sample is then annealed for 15 min at 1200 °C, which is below the desilication threshold ($\geq 1350$ °C). Subsequently, the sample is transferred back to the UHV chamber and degassed at 700 °C. The typical ARPES slice from samples annealed in Ar (figure 2.b) exhibits an apparent sharpening of the band structure as compared to annealing in UHV (figure 2.a). MDC comparison in figure 2.c, shows indeed a higher signal to noise contrast and sharper bands. The extracted bandwidth is $\Delta k = 0.12$ Å$^{-1}$ ± 0.02 Å$^{-1}$. Longer annealing in argon leads to growth of an additional carbon overlayer. As an alternative, we opted to help surface healing at high temperature by introducing a hydrocarbon gas (methane) in the UHV chamber. Like in chemical vapor deposition (CVD) of graphene, the hydrocarbon gas breaks at high temperatures and provide the surface with carbon atoms[34,35]. Here, after annealing at 950 °C in CH$_4$ pressure of 2×10$^{-6}$ mbar, ARPES displays a better band structure quality (figure 2.d to 2.f). Already after 1 hour, the bands are sharper than for Ar and have a bandwidth of $\Delta k = 0.11$ Å$^{-1}$ ± 0.02 Å$^{-1}$. The graphene bands become more intense after 13 hours of annealing with a slight decrease of the band width. Still, the XPS C 1s spectrum shows remaining defect contributions at the surface (figure 3). However, neither longer annealing nor increasing the annealing temperature up to 1000 °C could bring noticeable sharpening of the bands. Annealing at 1000 °C for more than one hour causes the appearance of a bilayer graphene signal, which gets stronger with annealing time. Thus, the best band structure quality is already achieved after 13 hours at 950 °C.

The microanalysis results are summarized in figure 4. LEEM measurements reveal the existence of two areas with bright and dark contrasts on the surface (figure 4.a). The estimated coverage of the bright area is around 90%. Similar features are observed during the analysis of the surface morphology using atomic force microscopy (AFM) (figure 4.b and 4.c). The tip-sample adhesion map (figure 4.c) matches the reverse-contrast of the LEEM data. The depicted vivid areas (figure 4.c) are located at the surface step edges, which are the initial growth sites of additional graphene layers. The evolution of the LEEM intensity as a function of incident energy (IV curve) from the bright area shows a single intensity dip located at E $\cong$ 2.1 eV, which is a clear signature of graphene decoupling[36,37]. This is supported by the corresponding µ-ARPES measurements. Figure 4.d shows a typical constant energy cut in the graphene band structure below the Dirac point. The presence of decoupled graphene is also verified by the detection of an intense graphene diffraction pattern and the suppression of reconstruction spots of ZLG[39] in the µ-LEED image (figure 4.e). The coincidence lattice between the graphene and the supporting SiC(0001) surface atomic structure is observed as a 13x13 periodicity near the (0,0) spot[38]. Besides, µ-LEED analysis shows the presence of additional spots having a $(\sqrt{3} \times \sqrt{3})R30°$ periodicity with respect to the SiC(1×1) (see inset in figure 4.e). The $(\sqrt{3} \times \sqrt{3})R30°$ superstructure has been known for the Si-rich SiC(0001) surface when additional Si atoms partially saturate the dangling bonds of the bare SiC(0001) surface and occupy the so-called T4 positions located above the carbon atoms in the first SiC bilayer[39]. Thus, the $(\sqrt{3} \times \sqrt{3})R30°$ here does not only add another hint that Bi atoms saturate dangling bonds of Si in an ordered configuration and decouples the graphene layer, but it also shows that Bi stabilizes the surface following the Si-rich SiC(0001) surface reconstruction. Still, since the LEED technique reveals only periodic structures, the configuration and arrangement of the Bi atoms on the surface is not completely clear. For Bi/Si(111), where a $(\sqrt{3} \times \sqrt{3})R30°$ reconstruction is observed, Bi atoms can arrange in monomers or in trimers[40]. Here, a trimer would result in a compact configuration of Bi atoms and Bi-Bi metallic bonding, which is not observed here (figure 1.f). Hence, a monomer phase is more plausible.

One reason for the persisting broad bands in the ARPES data could be the presence of unresolved two bands shifted from each other following a Rashba splitting, as expected by theoretical calculations[24]. In order to check the presence of spin-split bands we performed SARPES measurements. In Figure 5, the measured energy distribution curves recorded via the different spin–sensitive Mott-detectors show no difference. It is clear that the bands are not spin-polarized as expected by theory. These results are similar to when Bi is intercalated under graphene on Ni(111)[41]. They are imputed to the electronic configuration of Bi which, although heavy, lacks *d* electrons to hybridize with the π-bands and increase the SOC interaction[41,42]. In addition, here, the discrepancy with the theory is accentuated by the bonding nature of the intercalated Bi atoms with the SiC surface. Indeed, while in theory the intercalated Bi atoms stay in the metallic state[24], we find from the XPS analysis that Bi binds to surface Si atoms (figure 1.f). The 4f level of the intercalated Bi is shifted by 0.8 eV to higher binding energies in comparison to Bi deposit as a film on the SiC(0001) surface (metallic Bi), suggesting a charge transfer from Bi to SiC surface atoms and the formation of surface bismuth silicide or carbide. Although, bismuth silicide has not been extensively studied as a bulk compound[43], it has already been shown that Bi can bind to the Si(111) surface and form stable phases[40,44,45]. Meanwhile, the Bi 4f level from the bulk has a higher binding energy than for the surface. The corresponding chemical shift is comparable with $Bi_2O_3$, however, here no oxygen is detected. On the other hand, in the C 1s spectrum from the surface, the formation of stable phases of bismuth carbide that would contribute to the defect peak at higher binding energy is not excluded[46,47].

**Conclusion:**

We show that intercalation of epitaxial graphene on SiC(0001) can be performed by ion-implantation of the intercalant. Here, we used it for the intercalation of Bi which was not possible using state-of-the-art procedures. The spectro-microscopic characterization of the obtained system shows a successful decoupling of the graphene layer from the SiC(0001). Hence, ion-implantation complements the already available procedures to tune graphene properties on SiC(0001). Furthermore, using the desired intercalant in the form of an ion beam offers the possibility to focus and control the intercalation within a predefined nanostructure to fabricate graphene-based electronic devices.

**Acknowledgment:**

This work was supported by the German Research Foundation (DFG) in the framework of the Priority Program 1459 "Graphene". H.M.B. acknowledges funding from the German Research Foundation (DFG). We are indebted to the staff at BESSY II of the Helmholtz-Zentrum Berlin and would like to thank D. Marchenco for assistance.

**Figures**

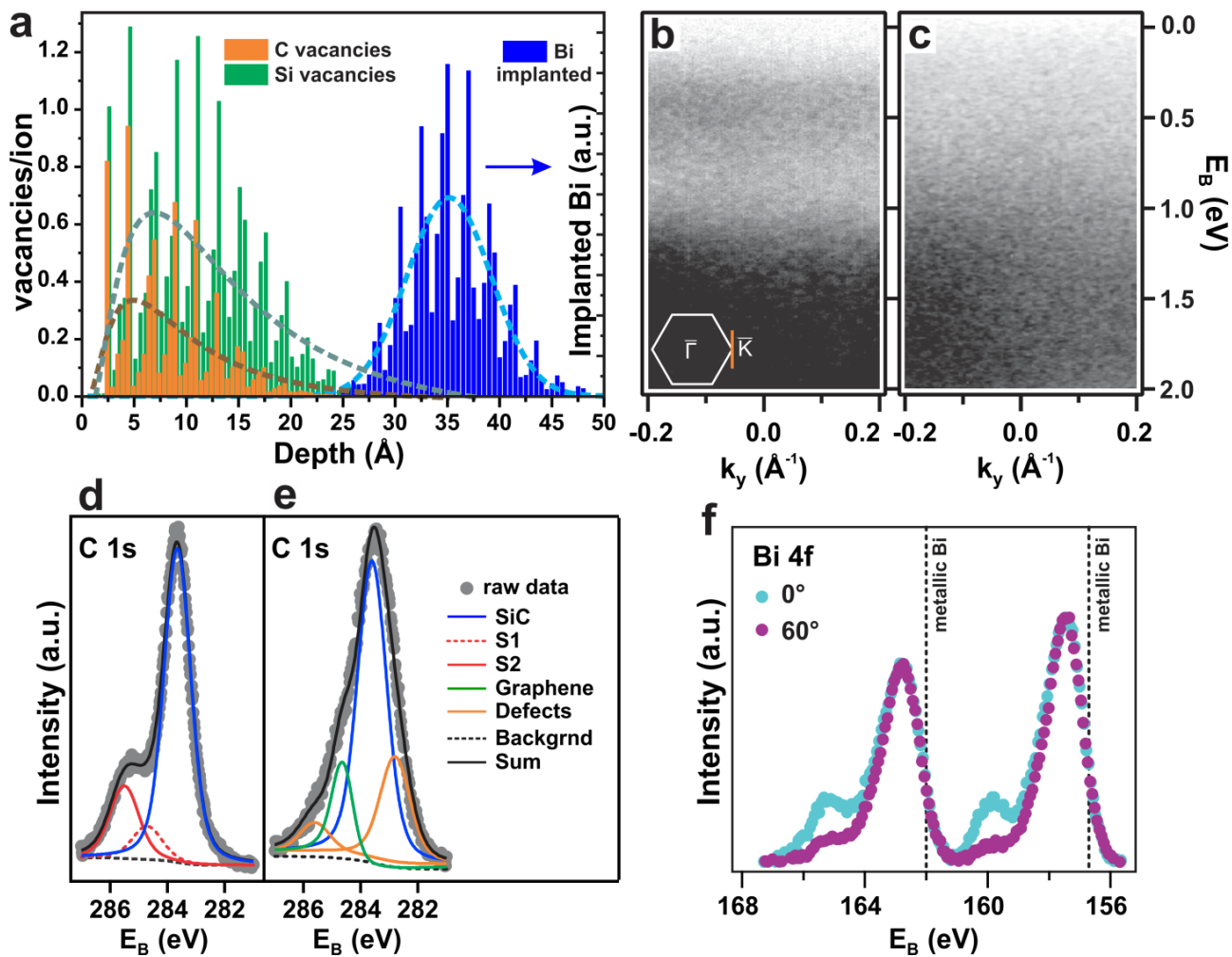

**Figure 1: a.** Depth profile of implanted Bi ions and created C and Si vacancies per impinging ion calculated using the SRIM program. The dashed lines are fits to the profiles. While the Bi profile presents a Gaussian distribution, the created vacancies are characterized by a log-normal distribution. **b.** and **c.** Experimental band structure recorded for a ZLG before and after Bi implantation, respectively. **d.** and **e.** XPS spectrum of the C 1s core level before and after Bi implantation, respectively. (Spectra are normalized to the 4f 7/2 peak height). **f.** Bi 4f core level spectra recorded at 0° (normal emission) and at 60°. The dashed lines indicate the position of the Bi 4f peak of a Bi thick film on SiC(0001).

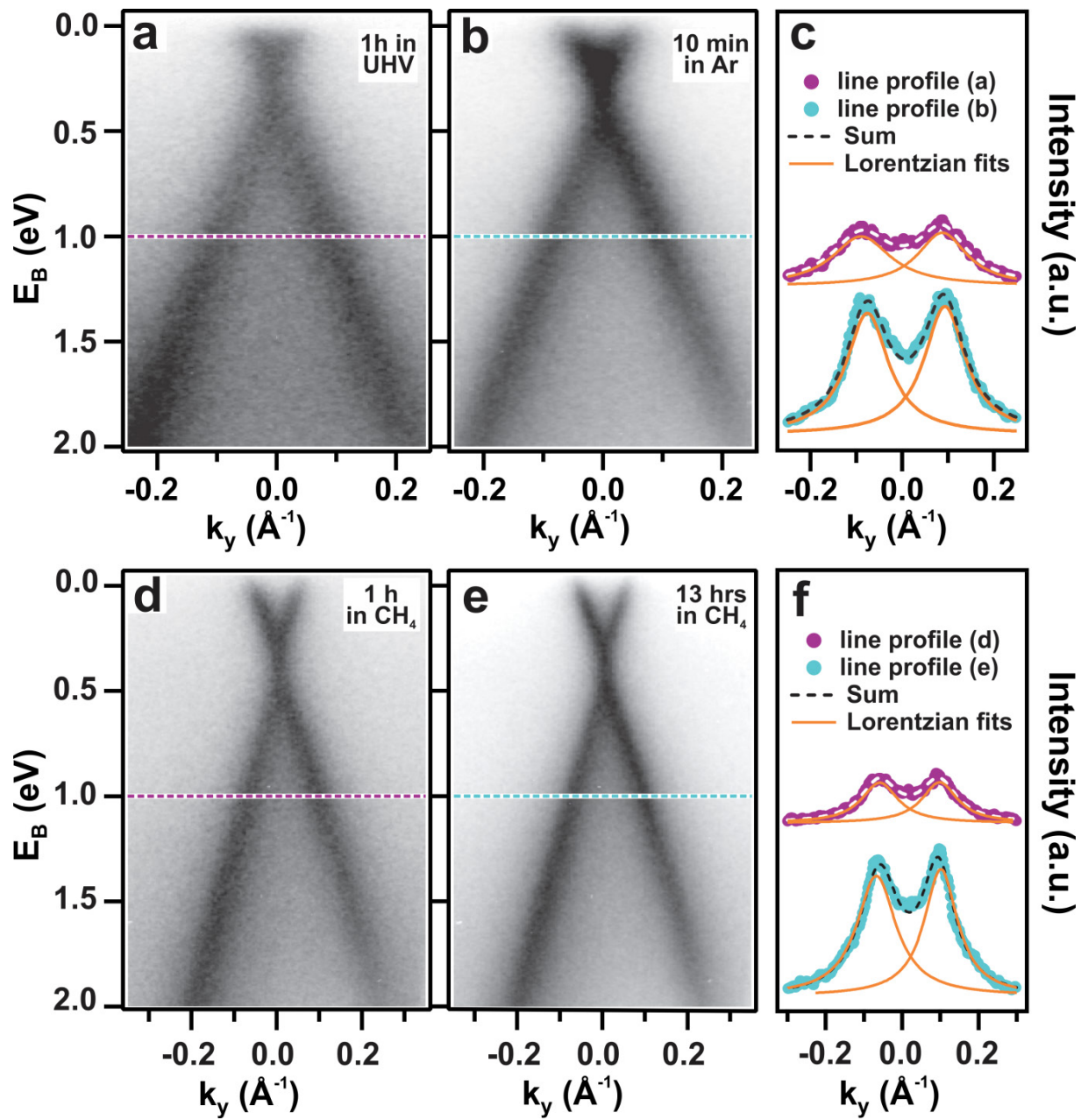

**Figure 2:** Experimental band structure of graphene on SiC(0001) after Bi implantation and annealing for **a.** 1 hour in UHV, **b.** 10 min in Argon atmosphere, **d.** 1 hour in Methane, and **e.** 13 hours in methane. The dashed lines indicate the position of the MDCs at $E_B = 1$ eV displayed in **c** and **f**.

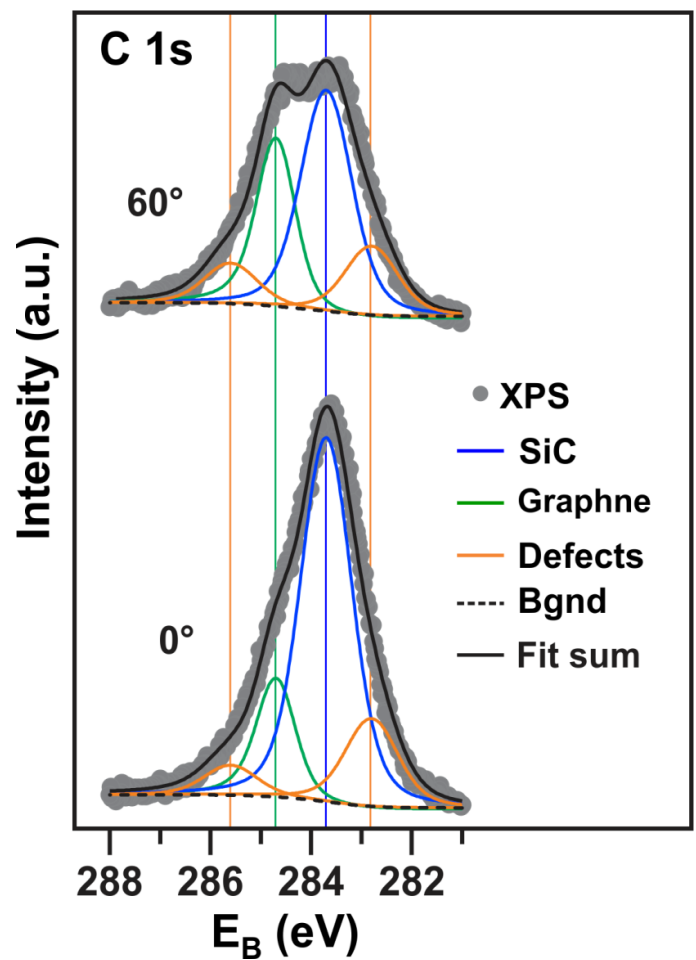

**Figure 3**: C 1s core level spectra recorded at 0° (normal emission) and at 60° from Bi intercalated graphene after 13 hours annealing in methane

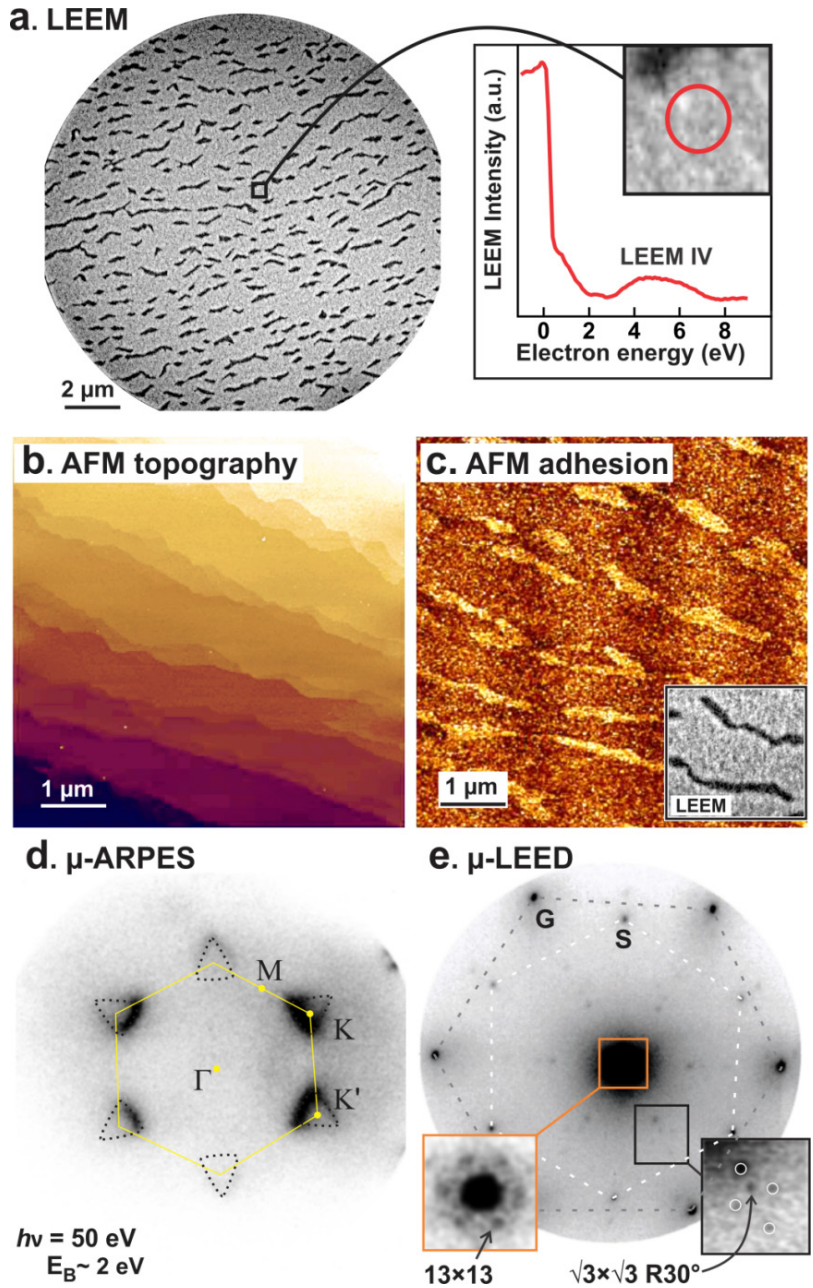

**Figure 4: Microanalysis of Bi intercalated graphene after annealing in CH$^4$ atmosphere for 13 hours. a.** LEEM micrograph together with a LEEM IV curve. The IV curve was recorded on the area delimited by the red circle in the bright zone indicated by the black square. The LEEM image was recorded for an electron energy of 3 eV. **b.** AFM topographical image of the surface and **c.** corresponding simultaneously recorded adhesion map. **d.** μ-ARPES constant energy slice. The yellow hexagon represents the 1$^{st}$ BZ of graphene. The black dashed triangles are guide to the eye. **e.** μ-LEED image of the surface recorded at 48 eV incident electron energy. The white and black dashed lines represent the LEED spots of graphene (G) and SiC (S), respectively. The white circles around the ($\sqrt{3} \times \sqrt{3}$)R30° indicate spots from the 13×13 coincidence pattern.

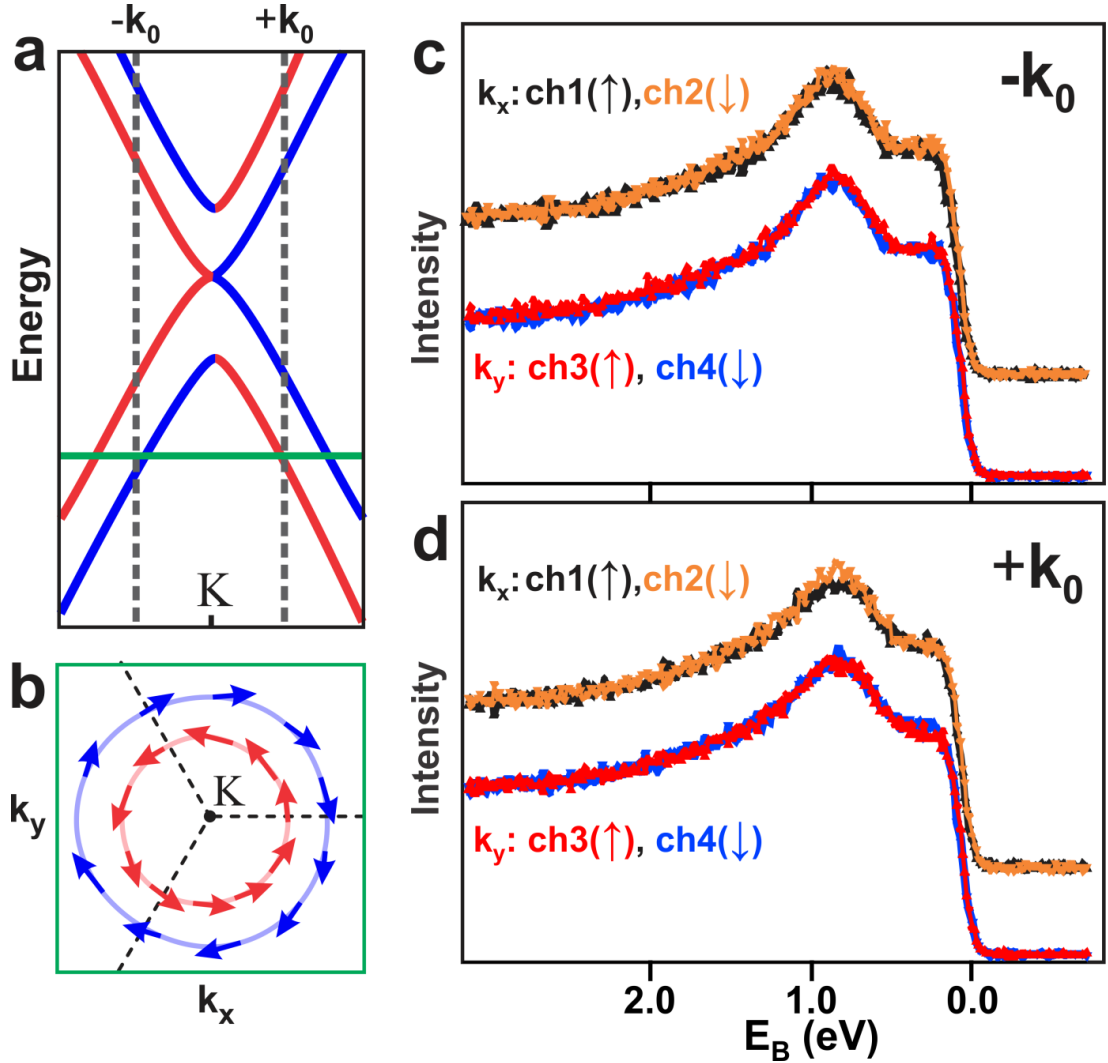

**Figure 5:** Schematic description of **a.** Rashba-split graphene band structure at the K point, and **b.** constant energy cut corresponding to the green line in a as described in ref [24]. The grey dashed lines show the momentum position where the spectra in c. and d. have been measured. **c.** and **d.** Spin-resolved ARPES for Bi-intercalated graphene measured at $k_0= 0.075$ Å$^{-1}$ on both sides of the K-point along $k_y$. ch1, ch2, ch3, and ch4 represent the channels of the Mott-detectors sensitive to electron spin-polarization along $k_x$ (ch1 and ch2) and $k_y$ (ch3 and ch4) in the momentum space (parallel to sample surface in direct space). The arrow sign indicates electronic spin up (↑) and spin down (↓).